\documentclass[12pt,titlepage]{article}

\usepackage{natbib}
\usepackage{graphicx}
\begin{document}

\title{Can we believe in high climate sensitivity?}
\author{J.D. Annan\thanks{email:
\texttt{jdannan@jamstec.go.jp}} and J.C. Hargreaves\\
Frontier Research Center for Global Change \\Japan Agency for Marine-Earth Science and Technology \\
3173-25 Showa-machi 
Kanazawa-ku \\ Yokohama, Kanagawa 236-0001, Japan }
 \maketitle

\bibliographystyle{../../packages/custom-bib/james}

%
%
%
%
%

%
%

%

%
%



%
%


\begin{abstract}
The climate response to anthropogenic forcing has long been one of the dominant uncertainties in predicting future climate change~\citep{ipcctar01}. Many observationally-based estimates of climate sensitivity ($S$) have been presented in recent years, with most of them assigning  significant probability to extremely high sensitivity, such as $P(S>6C)>5\%$. \\ However, closer examination reveals that these estimates are based on a number of implausible implicit assumptions. We explain why these estimates cannot be considered credible and therefore have no place in the decision-making process. In fact, when basic probability theory is applied and reasonable assumptions are made, much greater confidence in a moderate value for $S$ ($\simeq 2.5C$) is easily justified, with $S$ very unlikely to be as high as 4.5C.  
\end{abstract}

%
%

%


%
%

\section{Introduction}

The response of the climate system to anthropogenic forcing, traditionally  expressed as the equilibrium sensitivity (S) of the globally-averaged temperature to a doubling of the atmospheric concentration of CO2, has long been considered as having great significance in terms of our understanding of the climate system. A number of estimates have been presented over recent decades, perhaps the most famous being the statement of~\citet{charney79} that S was unlikely to lie outside the range of 1.5-4.5C, with that statement later formally presented as representing a probability somewhere in the range of 66-90\%~\citep{ipcctar01}. More recently, a proliferation of probabilistic estimates explicitly based on calculations using observational data have also been presented~\citep[eg][]{andsch01,grestorap02,forstosok02,hegcrohydfra06}. Many of these results suggest a worryingly high probability of high sensitivity, such as $P(S>6)>5\%$. The focus of this paper is to discuss how credible such estimates are.

To avoid possible misunderstandings, we establish at the outset that the notion of probability discussed here is the standard Bayesian paradigm of probability as the subjective degree of belief of the researcher in a proposition~\citep{bersmi94}. While this is not the only possible paradigm for the treatment of epistemic uncertainty in climate science~\citep[eg][]{kriegler05}, it appears to be the dominant one. The main (perhaps sole) reason for interest in such estimates is in order to support decision making, such as mitigation and adaptation strategies~\citep[eg][]{yohandsch04,meinshausen05}. In order for decisions made in the face of uncertainty to be rational, they must conform to the probability calculus~\citep{definetti74}, which {\em inter alia} mandates the application of Bayes' Theorem for updating beliefs in the light of evidence: $f(S|O)=f(O|S)f(S)/f(O)$, where $f(S|O)$ represents the posterior probability density function (pdf) for sensitivity S after taking account of a set of observations O, $f(S)$ represents the prior belief in the absence of these observations, $f(O|S)$ is the likelihood function which describes how the observations are probabilistically dependent on the sensitivity, and $f(O)$ is a normalisation factor. Beliefs which do not conform to the probability calculus are said to be incoherent, and are vulnerable to a Dutch Book argument. That is, is it possible to construct a sequence of decisions (bets), each one of which appears to be rational in the light of the stated beliefs, but which collectively ensure a loss under all possible outcomes. A simple example will be presented later.

In order to calculate $f(S|O)$, two inputs are required: the prior $f(S)$, and the likelihood $f(O|S)$ which depends on the observations which are used. We consider these two inputs in turn in the next two sections, and follow with some conclusions.

\section{The prior}

\subsection{``Ignorant'' priors}

If the posterior pdf is intended to represent the beliefs of the researcher concerning climate sensitivity after updating via a particular set of observations $O$, then the prior must logically represent their beliefs in the absence of these observations. Note that this does not actually require a chronological relationship between prior, observations, and posterior, although such a relationship may exist. Determining a suitable prior is potentially challenging given that in many or even most cases, the researcher is already aware of the broad implications of the data before the detailed quantitative analysis is undertaken. It is important to be aware of the risk of double-counting the data by accounting for it both in the prior and again through the likelihood, as committing this error would result in over-confident estimates. 

In an attempt to avoid that risk, researchers have often chosen to use a uniform prior, which is sometimes described as ''ignorant''. However, it must be recognised that in fact there can be no prior that actually represents a state of true ignorance regarding $S$. For example, any proper prior cannot avoid assigning a specific level of belief to the proposition that $S>6C$. Furthermore, the uniform priors which have been used represent beliefs that in our opinion are difficult to justify. The uniform prior U[0C,20C]~\citep{frabooket05} actually represents a  prior belief that S is ``likely'' (70\% probability) greater than 6C, with an expected value for S of 10C and a 50\% probability of exceeding this value. Even when truncated to U[0C,10C]~\citep{hegcrohydfra06}, such a uniform prior still represents the belief that $P(S>6C)=40\%$, and furthermore that $S$ is more than twice as likely to lie outside the conventional 1.5-4.5C ``likely'' range, as inside it. The notion that such priors can encapsulate the concept of ignorance may be superficially attractive but is surely not defensible in detail.

For illustration, we consider the ERBE data which were recently analysed by~\citet{forgre06}. This analysis has several attractive features which we discuss later. For now, it is sufficient to note that their analysis results in a likelihood which is Gaussian in feedback $L = 1/S$, which is broadly similar in shape to the marginal likelihood functions for sensitivity that have been obtained from a variety of investigations including those cited above. There are fundamental physical reasons for this which are well understood~\citep{hanruslac85}.

\begin{figure*}
\noindent\includegraphics[width=30pc]{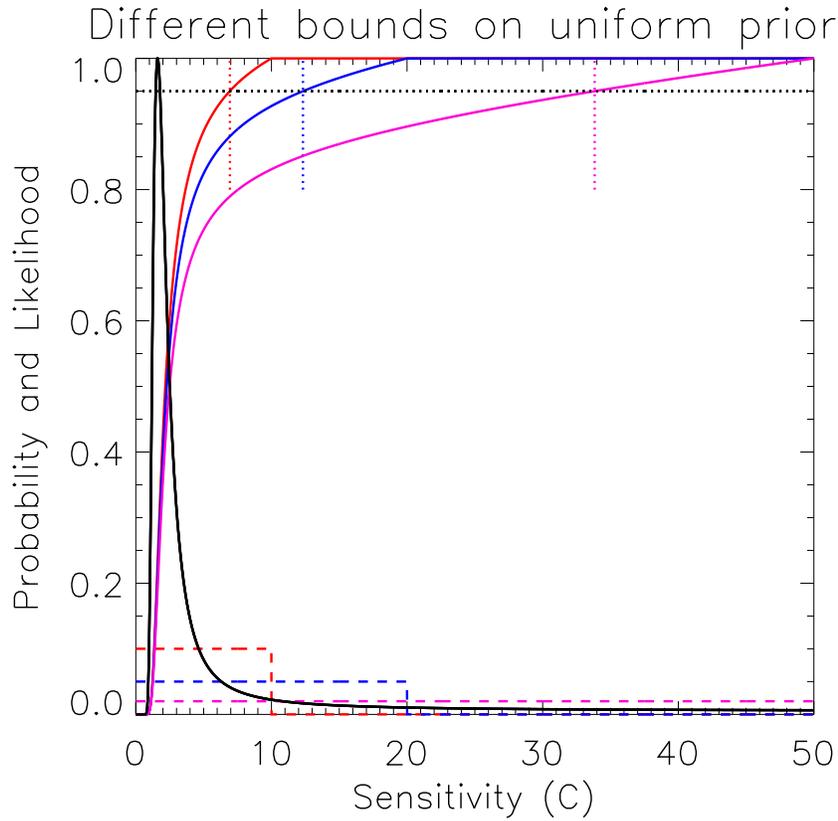}
\caption{Effect of using different bounds on a uniform prior. Solid black line indicates likelihood function of~\citet{forgre06}. Dashed coloured lines show the priors, solid coloured curves are cumulative  posterior pdfs and dotted lines indicate upper 95\% probability threshold. Red: U[0,10]~\citep{hegcrohydfra06},blue: U[0,20]~\citep{frabooket05}, magenta: U[0,50] }
\label{uniform}
\end{figure*}

A Gaussian in likelihood space has the unfortunate property that $f(O|L=0)$ is strictly greater than zero, and therefore for all large $S$, $f(O|S)$ is bounded below by a constant. Therefore, if the improper unbounded uniform prior is used, no proper posterior pdf results. This necessitates the selection of bounds on the uniform prior --- it would therefore be more appropriate to label it as ``a'' uniform prior --- and the results are strongly dependent on the upper bound selected. Figure~\ref{uniform} shows results obtained using three different uniform priors: U[0,10]~\citep{hegcrohydfra06}, U[0,20]~\citep{frabooket05} and an arbitrarily extended range of U[0,50]. In all cases, the likelihood is identical. That is, the difference in results here is entirely due to the choice of upper bound on the prior, rather than the observations. It would, we argue, be difficult to claim that any one of these choices was a more objective basis for decision making in preference to the other two, or other possible upper bounds. However, the posterior 95\% probability threshold is remarkably different between these three results, as is $P(S>6C)$. Choosing between these three (or other) alternatives could be expected to have strong implications for policy decisions.

It has even been suggested that a general method for probabilistic estimation can be established by choosing the prior to be uniform in the variable which is being estimated~\citep{frabooket05}. In addition to the need to choose bounds (for which no rationale has been presented), an even more fundamental problem with this approach is that it generates inconsistent results which do not conform to the probability calculus. Consider a single observation $X_o$  of an unknown variable $X$, which takes the value $X_o=2$ with an observational uncertainty of 0.5 (assumed to be the standard deviation of a Gaussian deviate). If we wish to estimate both $X$ and $Z=X^4$, then the proposal of~\citet{frabooket05} is that we should perform these estimations using a uniform prior in each variable in turn, which would generate the results that $P(X>3)=2.3\%$ but $P(Z>81)=P(X^4>3^4)=7.8\%$. Since both propositions are logically equivalent, assigning different probabilities to them is a clear indication of incoherence. At the risk of belabouring the point, a Dutch Book can be constructed by noting that the first probability implies the  acceptability of a bet which requires a stake of 0.97 (in units of utility) and which pays out 1 in return iff $X<3$, and the second probability implies a willingness to stake 0.07 on a bet which pays 1 iff $Z>81$. Collectively, the total stake is 1.04 and the return is only 1 irrespective of the actual outcome. It seems doubtful whether such an unconventional approach to probability can have a useful role to play in the decision making process.

Moreover, unless the set of data $O$ under explicit consideration actually include all of the evidence which might be considered relevant to the estimation of climate sensitivity, it is not even appropriate for the prior to represent ``ignorance'' at all. Instead, it should represent the background beliefs in the absence of $O$. Therefore, it seems clear there is no alternative but to attempt the task of selecting a prior that does in fact honestly represent the prior beliefs of the researcher --- that is, what they would believe in the absence of the data under examination. If confidence about the choice of such a prior is low, then a sensible response would be to test the sensitivity of the overall results to a range of reasonable choices, rather than abandon any attempt to undertake this estimation at all.

\subsection{Expert priors}

\begin{figure*}
\noindent\includegraphics[width=30pc]{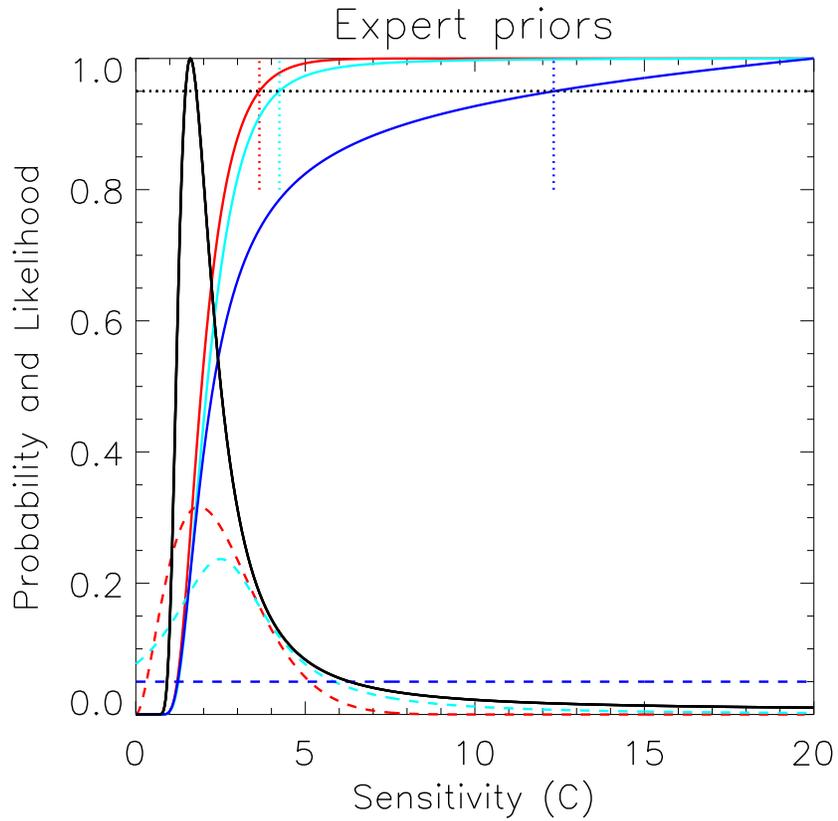}
\caption{Pdfs arising from expert priors. Solid black line indicates likelihood function of~\citet{forgre06}. Dashed coloured lines show the priors, solid coloured curves are cumulative  posterior pdfs and dotted lines indicate upper 95\% probability threshold. Red: Expert prior of~\citet{websok02}, cyan: extended high tail (see text) blue: U[0,20]}
\label{expert}
\end{figure*}

One way to attempt to formulate a more credible prior would be to look back through the literature, to see what climate scientists actually wrote prior to the analysis of modern data sets. After~\citet{arrhenius1896}'s early estimate of around 5C, all subsequent model-based estimates have been clearly lower~\citep{manwet67,hanrusrin83}, culminating in the ``likely'' range of 1.5--4.5C~\citep{charney79}. This is still well before any modern probabilistic analysis of the warming trend and much other data, and so could be considered a sensible basis for a  credible prior. Simple physically-based arguments also point towards a modest value as at least having higher probability than extremes. For example, the radiative forcing  effect of CO2 alone is estimated to be roughly 1C, with water vapour feedback doubling this to 2C~\citep{ipcctar01}. Cloud feedback is widely acknowledged to be highly uncertain, but a prior of U[0C,20C] requires the belief that not only it is ``very likely'' (90\%) positive, but furthermore likely to be large. We emphasise that we do not propose that such simplistic arguments can provide a precise estimate for $S$, or even that they justify the selection of a prior that completely prohibits high values. Rather, we merely use them to support our claim that the uniform priors which have been widely used represent an extreme viewpoint which cannot be reconciled with actual prior scientific  opinion. Returning to overtly subjective expert opinions, a composite expert prior has also been presented~\citep{websok02}, based on a survey of experts~\citep{morkei95}, which is also broadly consistent  with the long-held viewpoint that $S$ is likely to be moderate. It has already been shown by~\citet{forstosok02} that updating this expert prior with global temperature data from the 20th century results in greatly increased confidence in a moderate value for S. It is, however, hard to shake off the accusation that the experts who were surveyed in this case were aware of the recent warming rate, and had therefore already accounted for that data in their estimates. However, such an accusation can hardly remain credible if instead of using this historical temperature data, we consider the recent analysis of the ERBE data, which was only published more than 10 years after the survey (and note further that the raw observational data upon which the analysis was based entirely post-dates the Charney report so cannot possibly have influenced this assessment). We therefore update the expert prior with the likelihood function arising from the ERBE data, and present the results in Figure~\ref{expert}. The resulting 5-95\% posterior probability interval is 1.2-3.6C. This result is remarkably insensitive to reasonable changes in the prior. As a demonstration of this, we use an alternative prior with greatly exaggerated tails, also illustrated in Figure~\ref{expert}. This has the functional shape $f(S) \propto 1/((S-2.5)^2+3)$, truncated at 0C and 20C. This prior assigns substantially higher (and rather worrying) probabilities to extreme S, such as $P(S>6C)=15\%$ and $P(S>10C)=5\%$, and only 57\% probability to S lying in the traditional range of 1.5-4.5C. We suspect that if such an estimate had been presented in the Charney report, it would have been met with a mixture of widespread scepticism and alarm. Even in this case, however, the posterior 5-95\% probability range only widens to 1.3-4.2C. Such a result still represents a substantial improvement on all recent estimates. Furthermore, it can hardly be argued that this prior rules out high $S$ {\em a priori}, it merely assigns a substantial rather than extraordinarily high level of prior belief to such a hypothesis. For example, if the data actually indicated a strong likelihood for high sensitivity (say via a hypothetical likelihood function for radiative feedback given by L=N(0.4,0.1), which has maximum likelihood  at S=3.7/0.4=9.2C) then the posterior would have a 5-95\% probability range of 6.4-14.3C. Therefore, it is clear that the choice of such a prior in no way prevents the posterior from indicating a high probability of high sensitivity, if the data were to actually suggest this.

\section{The data}

As well as the choice of prior, the choice of observations $O$ is a crucial component in the analysis. Most researchers have only considered small sets of specific observations in isolation, such as globally-averaged temperature and forcing data~\citep{andsch01,grestorap02,forstosok02}, or the short-term cooling following a specific volcanic eruption~\citep{wigammsan05}. By returning to a uniform prior for the analysis of each new data set, it has proved possible to state that those specific data do not by themselves provide a good upper bound on $S$. However, this divide-and-conquer strategy cannot, by construction, generate probabilities that represent the beliefs of scientists who are aware of all (or even much) of the relevant data, and therefore has no direct value to decision-makers. If observations are not explicitly considered in the likelihood function, then they must be accounted for in the prior. More data can certainly be expected to reduce uncertainty, and it has been recently shown that substantial improvements can be expected from such an approach~\citep{annhar06,hegcrohydfra06}. We note that both of these analyses above were actually based on an underlying uniform prior, which implies that a more appropriate choice might have generated somewhat stronger results. There may be legitimate arguments about the conditional independence of various data from different analyses (especially when the analyses all require the use of a complex climate model which may introduce persistent biases), although there does not appear to be any meaningful discussion of this in the literature to date. The analysis of ERBE data seems particularly useful in this respect, since it is based on a direct regression analysis of satellite observations of radiation versus recent surface temperature data, and does not depend on climate models, (or even, say, the rate of heat diffusion into the ocean or the overall surface warming trend) in the generation of the likelihood function. Therefore, there can be little question over its independence from the prior estimates which we have discussed, or pdfs which have been published based on other data. Combining the ERBE analysis with the pdfs of~\citet{annhar06} and~\citet{hegcrohydfra06} noticeably sharpens their results, with an upper 95\% probability threshold of no more than 4C in each case.

\section{Conclusions}

If we are to act rationally based on probabilistic calculations, then it is essential to ensure that these decisions are based on credible analyses of the available evidence. By both choosing a uniform prior (which by construction assigns very high probability to high climate sensitivity), and also ignoring almost all data which would moderate this belief, researchers have generated a number of results which assign high probability to extremely high climate sensitivity. We have explained here why this approach is fundamentally unsound, and cannot be considered to plausibly represent the rational beliefs of informed climate scientists. If we use either one (let alone both) of (a) a plausible prior, even one which  assigns substantial (but not extraordinary) belief to high climate sensitivity, and (b) a somewhat more comprehensive analysis of multiple data sets, then the ``fat tail'' of high  sensitivity disappears, with an upper 95\% probability limit easily shown to lie close to 4C, and certainly well below 6C. These results are very robust with respect to realistic choices for the prior. Evidence arising from the analysis of observations can be considered either explicitly as part of the analysis, or implicitly in the prior, but cannot be arbitrarily ignored without disqualifying the resulting analysis from any claim to represent a credible belief. In the light of this analysis, it is difficult to see how a belief in a significant probability of very high climate sensitivity can be rationally sustained.

%
%


%
%
%
%
%
%
%
%




\end{document}